\documentclass{aa}\usepackage{graphics,epsf}
\newcommand{\lta}{\;
  \raise0.3ex\hbox{$<$\kern-0.75em\raise-1.1ex\hbox{$\sim$
  }}\;\hskip-2pt }
\newcommand{\gta}{\;
  \raise0.3ex\hbox{$>$\kern-0.75em\raise-1.1ex\hbox{$\sim$
  }}\;\hskip-2pt }
 \def\be{\begin{equation}}
 \def\ee{\end{equation}}
 \def\bea{\begin{eqnarray}}
 \def\eea{\end{eqnarray}}
\begin{document}
\thesaurus{06(06.13.1; 02.03.1; 02.13.1)}
\title{Dynamical variations of the differential rotation in the solar convection zone}
\author{Eurico Covas\thanks{e-mail: eoc@maths.qmw.ac.uk}\inst{1}
\and Reza Tavakol\thanks{e-mail: reza@maths.qmw.ac.uk}\inst{1}
\and David Moss\thanks{e-mail: moss@ma.man.ac.uk}\inst{2}
}
\institute{Astronomy Unit, School of Mathematical Sciences,
Queen Mary College, Mile End Road, London E1 4NS, UK
\and Department of Mathematics, The University, Manchester M13 9PL, UK
}
\offprints{\em E.\ Covas}
\authorrunning{Covas {\em et al.}}
\titlerunning{Dynamical variations in the solar convection zone}
\maketitle
\begin{abstract}
Recent analyses of helioseismological observations seem
to suggest the presence of two new phenomena connected with the dynamics
of the solar convective zone. Firstly,
there are present torsional oscillations
with periods of about 11 years,
which penetrate significantly into the solar convection zone
and secondly, oscillatory regimes exist near  the base
of the convection which are markedly different from those
observed near the top, having either significantly reduced
periods or being non-periodic.

Recently spatiotemporal fragmentation\,/\,bifurcation has been
proposed as a possible dynamical mechanism to
account for such observed multi-mode
behaviours in different parts of the
solar convection zone.
Evidence for this scenario was produced
in the context of an axisymmetric mean field
dynamo model operating in a spherical shell, with
a semi--open outer boundary condition and
a zero order angular velocity obtained by the
inversion of the MDI data,
in which the only nonlinearity was the action
of the Lorentz force of the
dynamo generated magnetic field on the solar
angular velocity.

Here we make a detailed study of the robustness
of this model with respect to plausible changes to its main
ingredients, including changes to the $\alpha$ and $\eta$ profiles as well as
the inclusion of a
nonlinear $\alpha$ quenching.
We find
that spatiotemporal fragmentation
is present in
this model for different choices of the rotation data and as
the details of the model are varied.
Taken together, these results give
strong support to the idea that spatiotemporal fragmentation
is likely to occur in general dynamo settings.
\end{abstract}


\section{Introduction}
Recent analyses of the helioseismological data, from both the
Michelson
Doppler Imager (MDI) instrument on board the SOHO
spacecraft (Howe et al.\ 2000a) and the Global Oscillation Network Group (GONG)
project (Antia \& Basu 2000)
have provided strong evidence to indicate that the previously observed
time variation of the differential rotation on the solar
surface -- the so called `torsional oscillations' with periods
of about 11 years (e.g.\ Howard \& LaBonte 1980; Snodgrass, Howard \& Webster 1985;
Kosovichev \& Schou 1997; Schou et al.\ 1998) -- penetrates into the
convection zone,
to a depth of at least 9 percent of the solar radius.
Torsional oscillations are
thought to be a consequence of
the nonlinear interactions between the magnetic
fields and the solar differential rotation. A number of attempts
have been made to model these oscillations. These include
modelling the dynamical feedback of the large scale magnetic field on the
turbulent Reynolds stresses that generate
the differential rotation in the convection zone
(the `nonlinear $\Lambda$--effect': Kitchatinov 1988; R\"udiger \& Kitchatinov 1990;
Kitchatinov et al.\ 1994;
K\"uker et al.\ 1996;
Kitchatinov \& Pipin 1998;
Kitchatinov et al.\ 1999), as well as models in which
the nonlinearity is through the direct action
of the azimuthal component of the Lorentz force of the
dynamo generated magnetic field on the solar
angular velocity (e.g.\ Schuessler 1981; Yoshimura 1981; Brandenburg \& Tuominen 1988;
Jennings 1993; Covas et al.\ 2000a,b; Durney 2000).

Further studies of these data have produced intriguing, but rather
contradictory results. Howe et al.\ (2000b) find evidence for the presence of
such oscillations around the tachocline near the bottom of the convection
zone, but with markedly shorter periods of about $1.3$ years.
On the other hand, Antia \& Basu (2000) do not find such oscillations
at the bottom of the convective zone.
Whatever the true dynamical behaviour at these lower levels may be,
the crucial point about these recent results is that
they both seem to indicate the possibility that
the variations in the differential rotation
can have different dynamical modes of behaviour
at different depths in the solar
convection zone: oscillations with a very different period
in the former case and non-periodic behaviour in the latter.

Clearly, further observations are required
to clarify this situation. However, whatever the outcome of such observations,
it is of interest to ask
whether such different variations
can in principle occur in different spatial locations in the convection zone and,
if so, what could be the possible mechanism(s) for their production.
This is of particular interest
given the inevitable errors
in helioseismological inversions,
especially as depth increases.

Recently,
{\it spatiotemporal fragmentation/bifurcation}
has been
proposed as a possible dynamical mechanism to
account for the observed multi-mode
behaviour at different parts of the
solar convection zone (Covas et al.\ 2000b) (hereafter CTM).
This occurs when
dynamical regimes
which possess different temporal behaviours coexist at
different spatial locations,
at {\it given} values of the control parameters of the system.
The crucial point is that these
different dynamical modes of behaviour can occur
without requiring changes in the
parameters of the model,
in contrast to the usual
temporal bifurcations which result in
identical temporal behaviour at each spatial
point, and which occur subsequent to changes in the model parameters.
Also, as we shall see below, spatiotemporal fragmentation/bifurcation
is a dynamical mechanism, the occurrence of which does not depend upon
the detailed physics at different spatial locations.

Evidence for the occurrence of this
mechanism was produced in CTM in the context of
a two dimensional axisymmetric mean field
dynamo model operating in a spherical shell, with
a semi--open outer boundary condition,
in which the only nonlinearity  is the action
of the azimuthal component of the Lorentz force of the
dynamo generated magnetic field on the solar
angular velocity. The
zero order angular velocity was
chosen to be consistent with the most recent helioseismological (MDI) data.
Despite the success of this model in producing
spatiotemporal fragmentation, a
number of important questions
remain. Firstly, there are error bars
due to the nature of the observational data as well as
inversion schemes used.
Secondly, the model used by CTM is approximate
and includes
many simplifying assumptions.
As a result, to be certain that
spatiotemporal fragmentation
can in fact be produced as a result of
such nonlinear interactions, independently of the
details of the model employed,
it is necessary that it is robust, i.e. that it can  produce
such behaviour independently of these details.

The aims of this paper are twofold. Firstly, we
make a systematic study of
of this mechanism by making a comparison
of the cases where the zero order rotational velocity are given by
the inversions of the MDI and GONG data respectively.
Given the detailed differences between these rotation profiles,
this amounts to studying the robustness of the
mechanism with respect to small changes in the
zero order rotation profile.

Secondly,
we study the
robustness of this model (in producing
spatiotemporal fragmentation) with respect to
a number of plausible changes to
its main ingredients, and
demonstrate that the occurrence of this mechanism
is not dependent upon the details of our model.

We show that in addition to producing butterfly
diagrams which are in qualitative agreement
with the observations as well as displaying
torsional oscillations that penetrate
into the convection zone, as recently  observed
by Howe et al.\ (2000a) and Antia \& Basu (2000),
and studied by Covas et al.\ (2000a),
the model can produce qualitatively
different forms of spatiotemporal fragmentary behaviours,
which could in principle account for either of the contradictory types of dynamical behaviour
observed by Howe et al.\ (2000b) and Antia \& Basu (2000) at the bottom of the convection zone.

The structure of the paper is as follows.
In the next section we outline our model.
Section 3 contains our detailed results for both MDI and GONG
data. In section 4 we study the robustness of spatiotemporal
fragmentation with respect to various changes in the details
of our model and finally section 5
contains our conclusions.
\section{The model}
We shall assume that the gross features of the
large scale solar magnetic field
can be described by a mean field dynamo
model, with the standard equation
\begin{equation}
\frac{\partial{\bf B}}{\partial t}=\nabla\times({\bf u}\times {\bf B}+\alpha{\bf
B}-\eta\nabla\times{\bf B}),
\label{mfe}
\end{equation}
where $\vec{B}$ and $\vec{u}$ are the mean magnetic field and the mean
velocity respectively. The quantities $\alpha$ (the $\alpha$ effect) and the
turbulent magnetic diffusivity $\eta_t$, appear in the process of the
parameterisation of the second order correlations between the velocity and
magnetic field fluctuations ($\vec{u}'$ and $\vec{B}'$).
Here ${\bf u}=v\mathbf{\hat{\vec{\phi}}}-\frac{1}{2}\nabla\eta$,
the term proportional to
$\nabla\eta$ represents the effects of turbulent diamagnetism,
and the velocity field is taken to be of the form $
v=v_0+v'$,
where $v_0=\Omega_0 r \sin\theta$, $\Omega_0$ is a prescribed
underlying rotation law and the component $v'$ satisfies
\begin{equation}
\frac{\partial v'}{\partial t}=\frac{(\nabla\times{\bf B})\times{\bf B}}{\mu_0\rho
r \sin\theta} . \mathbf{\hat{\vec{\phi}}}  + \nu D^2 v',
\label{NS}
\end{equation}
where $D^2$ is the operator
$\frac{\partial^2}{\partial r^2}+\frac{2}{r}\frac{\partial}{\partial r}+\frac{1}
{r^2\sin\theta}(\frac{\partial}{\partial\theta}(\sin\theta\frac{\partial}{\partial
\theta})-\frac{1}{\sin\theta})$  and $\mu_0$ is the induction constant.
The assumption of axisymmetry allows the field ${\bf B}$ to be split simply
into toroidal and poloidal parts,
${\bf B}={\bf B}_T+{\bf B}_P = B\hat{\vec{\phi}} +\nabla\times A\hat{\vec{\phi}}$,
and Eq.\ (\ref{mfe}) then yields two scalar equations for $A$ and $B$.
Nondimensionalizing in terms of the solar radius $R$ and time $R^2/\eta_0$,
where $\eta_0$ is the maximum value of $\eta$, and
putting $\Omega=\Omega^*\tilde\Omega$, $\alpha=\alpha_0\tilde\alpha$,
$\eta=\eta_0\tilde\eta$, ${\bf B}=B_0\tilde{\bf B}$ and $v'= \Omega^* R\tilde v'$,
results in a system of equations for $A,B$
\bea
\label{A-eq}
\frac{\partial A}{\partial \tau} &=& w_rB_\theta-w_\theta B_r+R_\alpha\tilde\alpha B
+\tilde\eta D^2A, \\
\label{B-eq}
\frac{\partial B}{\partial \tau} &=&R_\omega r\sin\theta{\bf B} . \nabla\Omega -
\frac{1}{r}\frac{\partial}{\partial r}(rw_r B) \\
&-& \frac{1}{r}\frac{\partial}{\partial
\theta}(w_\theta B)+\frac{1}{r^2\sin\theta}\frac{\partial\tilde\eta}{\partial
\theta}(B\sin\theta) \nonumber \\
&+& R_\alpha \left (-\tilde\alpha D^2 \alpha +
\frac{\partial \tilde\alpha}{\partial r} B_\phi -
\frac{1}{r} \frac{\partial \tilde\alpha}{\partial \theta} B_r \right ) \nonumber \\
&+& \frac{1}{r}\frac{\partial \tilde\eta}{\partial r}\frac{\partial
}{\partial r}(rB)+\tilde\eta D^2 B, \nonumber
        \eea
where ${\bf w}=-\frac{1}{2}\nabla\eta$.
The
dynamo parameters are $R_\alpha=\alpha_0R/\eta_0$, $R_\omega=\Omega^*R^2/\eta_0$,
$P_r=\nu_0/\eta_0$, and $\tilde\eta=\eta/\eta_0$, where $\Omega^*$
is the solar surface equatorial angular velocity.
Here
$\nu_0$ and $\eta_0$ are the values of the turbulent magnetic
diffusivity and viscosity respectively
and $P_r$ is the turbulent Prandtl number.
The density $\rho$ is assumed to be uniform.

For our inner boundary conditions we chose $B=0$ on $r=r_0$
(which enforces angular momentum conservation in the dynamo region, as
discussed in Moss \& Brooke 2000).
At the outer boundary, we used an open
boundary condition $\partial B/\partial r = 0$ on $B$, and
vacuum boundary conditions for ${\bf B}_P$ (as in CTM). The motivation for
the former condition
is that the surface boundary condition is ill-defined, and there is some
evidence that the more usual $B=0$ condition may be inadequate.
This issue has recently been discussed at length by Kitchatinov, Mazur \& Jardine
(2000), who derive `non-vacuum' boundary conditions on both
$B$ and ${\bf B}_P$.

Equations (\ref{NS}), (\ref{A-eq}) and (\ref{B-eq}) were solved using the code
described in Moss \& Brooke (2000) (see also
Covas et al.\ 2000a,b) together with the above boundary conditions,
over the range $r_0\leq r\leq1$, $0\leq\theta\leq \pi$.
We set  $r_0=0.64$; with
the solar convection zone proper being thought to occupy the region $r \gta 0.7$,
the region $r_0 \leq r \lta 0.7$ can be thought of as an overshoot
region/tachocline.
In the following simulations we used a mesh resolution of $61 \times 101$
points, uniformly distributed in radius and latitude respectively.

The model employed in CTM had the following ingredients.
In the interval $0.64\leq r \leq 1$,  $\Omega_0$ was given
by the inversion of the MDI data
obtained from 1996 to 1999 by Howe et al.\ (2000a).
Here, in addition to this, we shall for comparison
also use the
$\Omega_0$ given
by the inversion of the GONG data
obtained from 1995 to 1999 by Howe et al.\ (2000b).
The form of
$\alpha$ was taken as
\begin{equation}
\label{tildealpha}
\tilde\alpha=\alpha_r(r)f(\theta),
\end{equation}
where
$f(\theta)=\sin^4\theta\cos\theta$
(cf.\ R\"udiger \& Brandenburg 1995).
The model possessed radial dependence in both
$\alpha$ and the turbulent diffusion coefficient $\eta$:
the $\alpha$ profile
was chosen by setting
$\alpha_r=1$ for $0.7 \leq r \leq 0.8$
with cubic interpolation to zero at $r=r_0$ and $r=1$,
with the convention that $\alpha_r>0$
and $R_\alpha < 0$.
The $\eta$ profile was chosen
in order to take into account the
likely decrease of $\eta$
in the overshoot region, by allowing
a simple
linear decrease in $\tilde\eta$
from $\tilde\eta=1$ at $r=0.8$
to $\tilde\eta=0.5$ in $r<0.7$.

In the following sections we
shall, in addition to these particular profiles,
also consider variations to them in order
to study the robustness of the model (in producing spatiotemporal fragmentation)
with respect to such changes.
Also for the sake of comparison, we shall
use the interpolation on the GONG data for the rotation law
as well as allowing other changes.
\section{Torsional oscillations and
fragmentation using the MDI and GONG data sets}
\label{tos}
\begin{figure}[!htb]
\centerline{\def\epsfsize#1#2{0.5#1}\epsffile{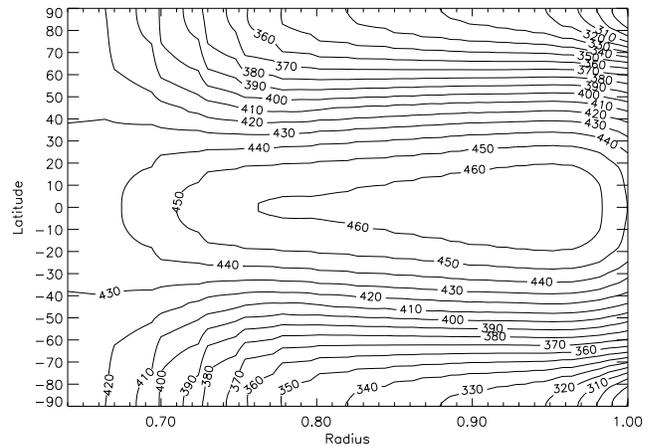}}
\caption{\label{gong} Isolines of the time average of the angular velocity of the
solar rotation, obtained by inversion techniques
using the GONG data (Howe {\em et al.} 2000b).
Contours are labelled in units of nHz.}
\end{figure}

\begin{figure}[!htb]
\centerline{\def\epsfsize#1#2{0.5#1}\epsffile{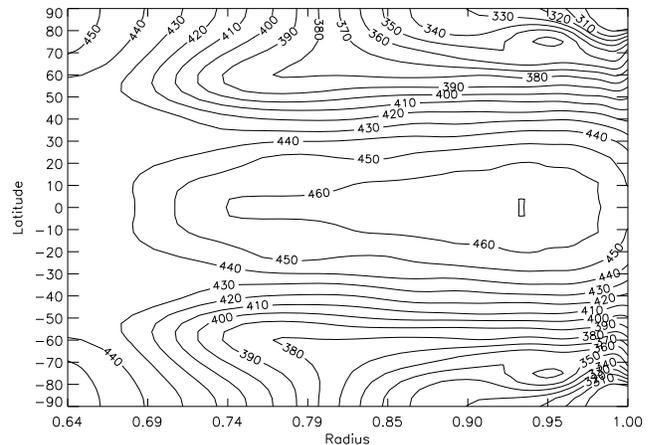}}
\caption{\label{mdi} Isolines of the time average of the angular velocity of the
solar rotation, obtained by inversion techniques
from the MDI data (Howe {\em et al.} 2000a).
Contours are labelled in units of nHz.}
\end{figure}

In this section, we
study the presence of spatiotemporal fragmentation
and torsional oscillations, using the zero order angular velocity
obtained by the inversion of the GONG data
(see Fig.\ \ref{gong}).
This allows a detailed
comparison to be made with the results of CTM which employed the
corresponding MDI data (see Fig.\ \ref{mdi}), including
the magnetic field
structure and strengths as well as the  nature of torsional oscillations,
as a function of
depth and latitude.

To begin with, we calibrated our model in each case
so that near marginal excitation
the cycle period was about 22 years.
This determined $R_\omega$ to be $44000$,
corresponding to
$\eta_0\approx 3.4 \times 10^{11}$ cm$^2$ sec$^{-1}$,
given the known values of $\Omega^*$ and $R$.

In both cases, the first solutions to be excited
in the linear theory
are limit cycles with odd (dipolar) parity
with respect to the equator, with
marginal dynamo numbers
$R_\alpha \approx -3.05$ and $-3.11$ respectively for the GONG and
MDI
data.
The even parity (quadrupolar) solutions
are also excited at similar marginal
dynamo numbers of $R_\alpha \approx -3.20$ and
$-3.22$, respectively. We note that these marginal dynamo numbers
are slightly different
from those reported in CTM as we have used here
$f(\theta)=\sin^4\theta\cos\theta$
for the $\theta$--dependence of the $\alpha$ effect, whilst CTM
used the prescription $f(\theta)=\sin^2\theta\cos\theta$.

Also, to be consistent with CTM, we chose the value of
the Prandtl number in this section to be $P_r = 1.0$.
For the parameter range that we investigated, the even parity solutions
can be nonlinearly stable. Given that the Sun is observed to be close to
an odd (dipolar) parity state, and that previous experience shows that small
changes in the physical model can cause a change between odd and even parities
in the stable nonlinear solution, we chose to impose dipolar parity on our
solutions, effectively solving the equations in one quadrant,
$0\leq\theta\leq\pi/2$, and imposing appropriate boundary conditions
at the equator $\theta=\pi/2$.

\begin{figure}[!htb]
\centerline{\def\epsfsize#1#2{0.38#1}\epsffile{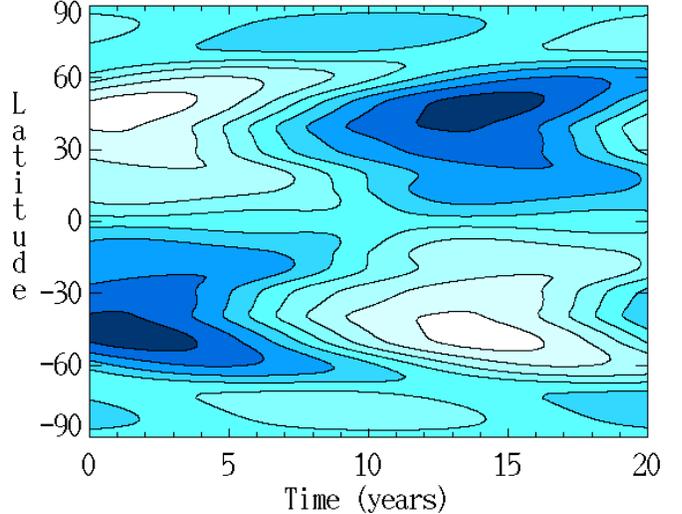}}
\caption{\label{dp_r=0.64_Calph=-11.0_iangalp=11_p=-1.0_Comega=44000_ialp=3_mesh=061x101_pr=1.0_ivac=2_eta=0_NS=1_rot=MDI2_rho=1.butterfly_bp}
Butterfly diagram of the toroidal component of the
magnetic field $\vec{B}$ at fractional radius $r=0.95$
for the MDI data.
Dark and light shades correspond to positive and negative values of
$B_\phi$ respectively.
Parameter values
are $R_\alpha=-11.0$, $P_r=1.0$ and ${R_{\omega}}=44000$.
}
\end{figure}

\begin{figure}[!htb]
\centerline{\def\epsfsize#1#2{0.38#1}\epsffile{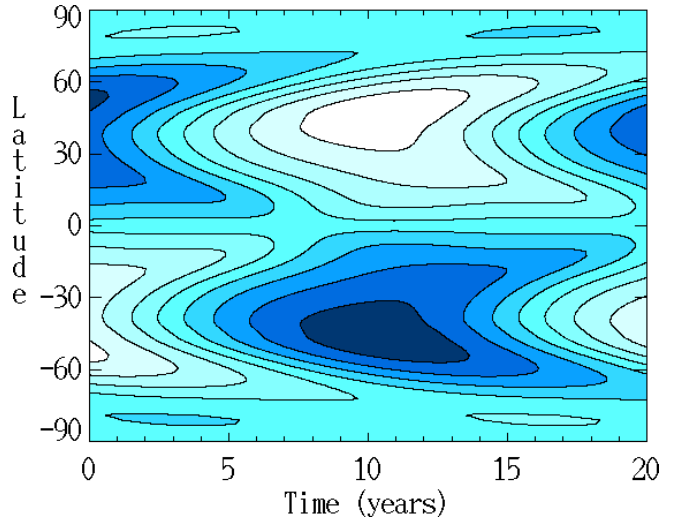}}
\caption{\label{dp_r=0.64_Calph=-9.0_iangalp=11_p=-1.0_Comega=44000_ialp=3_mesh=061x101_pr=1.0_ivac=2_eta=0_NS=1_rot=GONG_rho=1.butterfly_bp}
Butterfly diagram of the toroidal component of the
magnetic field $\vec{B}$ at fractional radius $r=0.95$
for the GONG data.
Dark and light shades correspond to positive and negative values of
$B_\phi$ respectively.
Parameter values
are $R_{\alpha}=-9.0$, $P_r=1.0$ and ${R_{\omega}}=44000$.
}
\end{figure}

With these parameter values, we found that this model,
with the underlying zero order angular velocity
chosen to be consistent with either the MDI or the
GONG data,
is capable of
producing butterfly diagrams which are in qualitative agreement
with the observations, as can be seen in
Figs.\
\ref{dp_r=0.64_Calph=-11.0_iangalp=11_p=-1.0_Comega=44000_ialp=3_mesh=061x101_pr=1.0_ivac=2_eta=0_NS=1_rot=MDI2_rho=1.butterfly_bp}
and
\ref{dp_r=0.64_Calph=-9.0_iangalp=11_p=-1.0_Comega=44000_ialp=3_mesh=061x101_pr=1.0_ivac=2_eta=0_NS=1_rot=GONG_rho=1.butterfly_bp}.
We note that the butterfly diagrams in the GONG case tend to be smoother and to
have weaker
polar branches, presumably because the MDI inversion is respectively less smooth and has a
distinctive
polar jet close to $r=0.95 R_{\odot}$.

\begin{figure}[!htb]
\centerline{\def\epsfsize#1#2{0.38#1}\epsffile{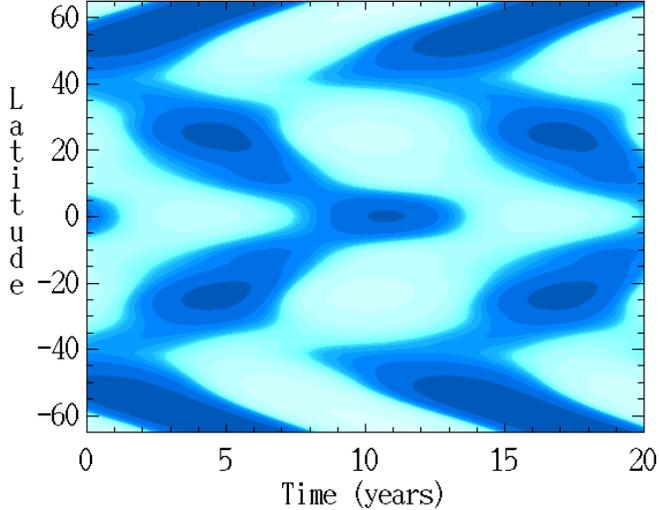}}
\caption{\label{dp_r=0.64_Calph=-11.0_iangalp=11_p=-1.0_Comega=44000_ialp=3_mesh=061x101_pr=1.0_ivac=2_eta=0_NS=1_rot=MDI2_rho=1.velocity.R=0.95}
Variation of the perturbation to the zero order rotation rate in latitude and time,
revealing the migrating banded zonal flows, taken at fractional radius $r=0.95$,
with the MDI data.
Parameter values
are as in Fig.\ {\protect
\ref{dp_r=0.64_Calph=-11.0_iangalp=11_p=-1.0_Comega=44000_ialp=3_mesh=061x101_pr=1.0_ivac=2_eta=0_NS=1_rot=MDI2_rho=1.butterfly_bp}}
}
\end{figure}

\begin{figure}[!htb]
\centerline{\def\epsfsize#1#2{0.38#1}\epsffile{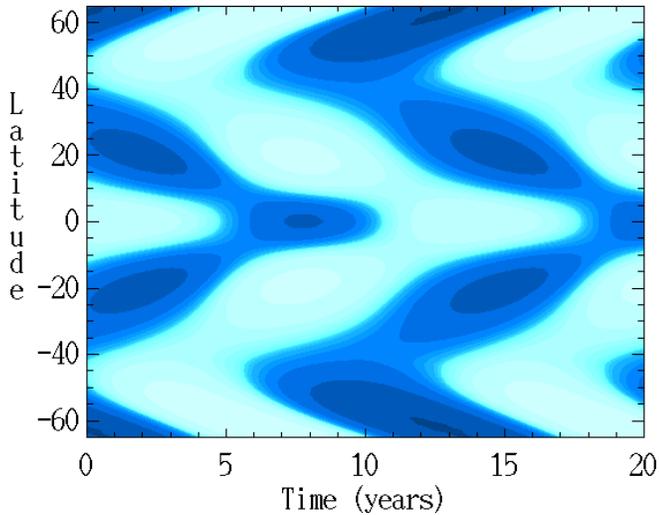}}
\caption{\label{dp_r=0.64_Calph=-9.0_iangalp=11_p=-1.0_Comega=44000_ialp=3_mesh=061x101_pr=1.0_ivac=2_eta=0_NS=1_rot=GONG_rho=1.velocity.R=0.95}
Variation of the perturbation to the zero order rotation rate in latitude and time,
revealing the
migrating banded zonal flows, taken at fractional radius $r=0.95$
with the GONG data.
Parameter values are as in Fig.\ {\protect
\ref{dp_r=0.64_Calph=-9.0_iangalp=11_p=-1.0_Comega=44000_ialp=3_mesh=061x101_pr=1.0_ivac=2_eta=0_NS=1_rot=GONG_rho=1.butterfly_bp}}.}
\end{figure}

The model can also produce torsional oscillations
in both cases (see Figs.\
\ref{dp_r=0.64_Calph=-11.0_iangalp=11_p=-1.0_Comega=44000_ialp=3_mesh=061x101_pr=1.0_ivac=2_eta=0_NS=1_rot=MDI2_rho=1.velocity.R=0.95}
and
\ref{dp_r=0.64_Calph=-9.0_iangalp=11_p=-1.0_Comega=44000_ialp=3_mesh=061x101_pr=1.0_ivac=2_eta=0_NS=1_rot=GONG_rho=1.velocity.R=0.95}),
that penetrate into the convection zone, in a manner similar to those deduced from
recent helioseismological data (Howe  et al.\ 2000a; Antia \& Basu 2000) and studied in
Covas et al.\ (2000a,b).

\begin{figure}[!htb]
\centerline{\def\epsfsize#1#2{0.38#1}\epsffile{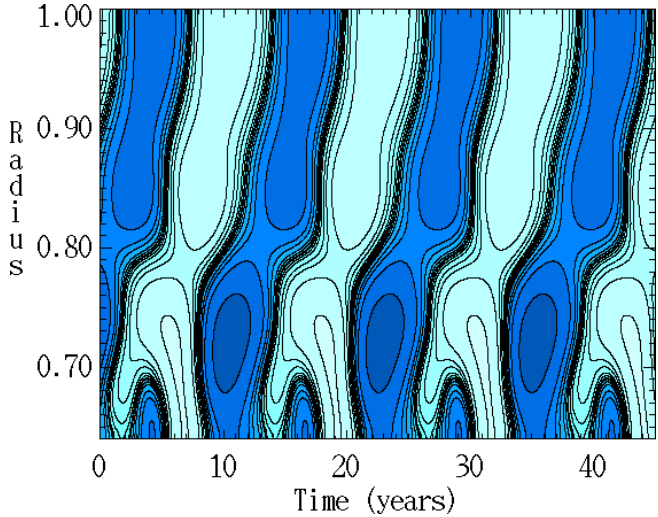}}
\caption{\label{dp_r=0.64_Calph=-11.0_iangalp=11_p=-1.0_Comega=44000_ialp=3_mesh=061x101_pr=1.0_ivac=2_eta=0_NS=1_rot=MDI2_rho=1.velocity_radial_latitude=30}
Radial contours of the angular velocity residuals $\delta \Omega$
as a function of time for a cut at latitude $30^{\circ}$,
with the MDI data.
Parameter values
are as in Fig.\ {\protect
\ref{dp_r=0.64_Calph=-11.0_iangalp=11_p=-1.0_Comega=44000_ialp=3_mesh=061x101_pr=1.0_ivac=2_eta=0_NS=1_rot=MDI2_rho=1.butterfly_bp}}.
}
\end{figure}

\begin{figure}[!htb]
\centerline{\def\epsfsize#1#2{0.38#1}\epsffile{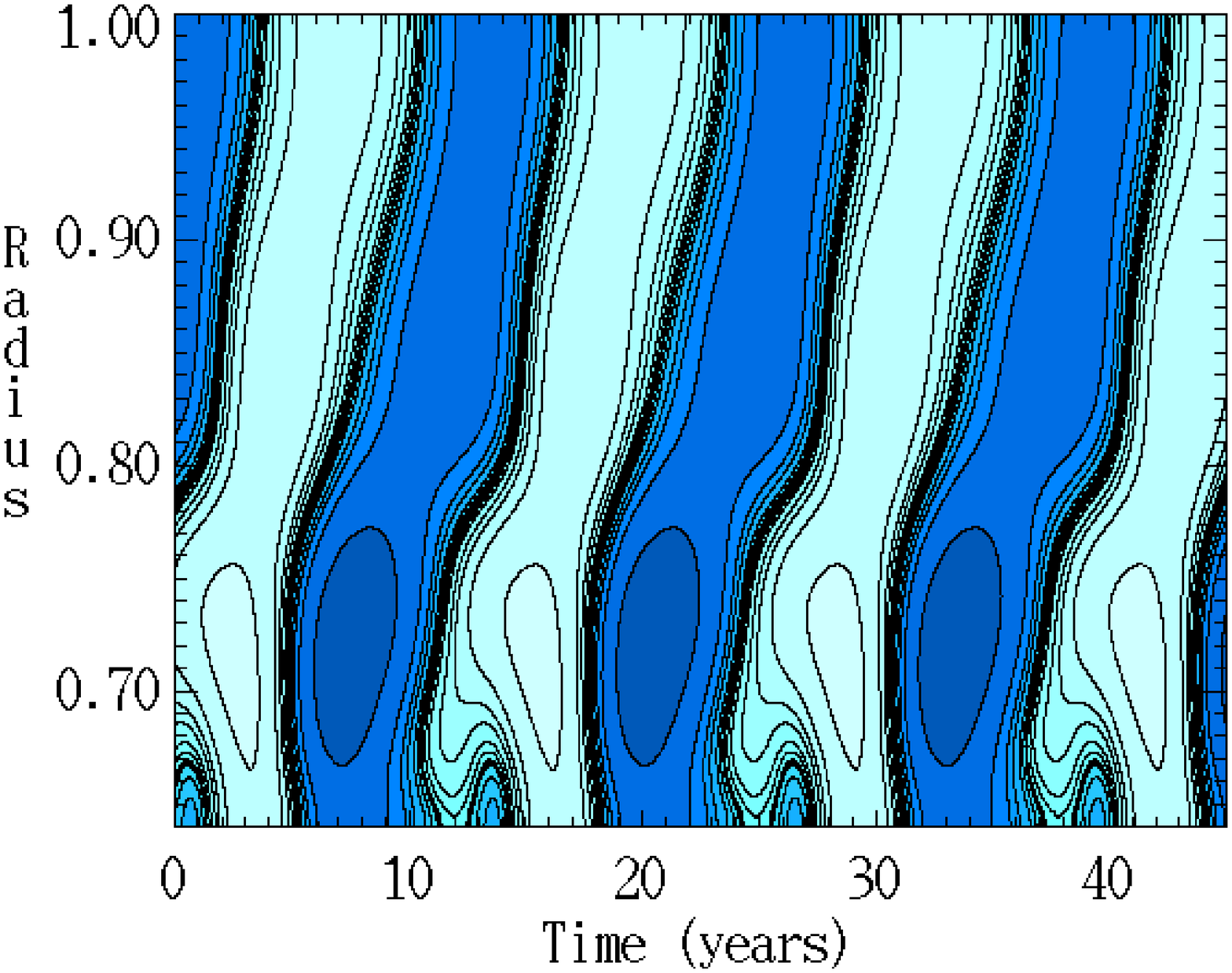}}
\caption{\label{dp_r=0.64_Calph=-9.0_iangalp=11_p=-1.0_Comega=44000_ialp=3_mesh=061x101_pr=1.0_ivac=2_eta=0_NS=1_rot=GONG_rho=1.velocity_radial_latitude=30}
Radial contours of the angular velocity residuals $\delta \Omega$
as a function of time for a cut at latitude $30^{\circ}$,
for the GONG data.
Parameter values are as in Fig.\ {\protect
\ref{dp_r=0.64_Calph=-9.0_iangalp=11_p=-1.0_Comega=44000_ialp=3_mesh=061x101_pr=1.0_ivac=2_eta=0_NS=1_rot=GONG_rho=1.butterfly_bp}}.}
\end{figure}

\begin{figure}[!htb]
\centerline{\def\epsfsize#1#2{0.42#1}\epsffile{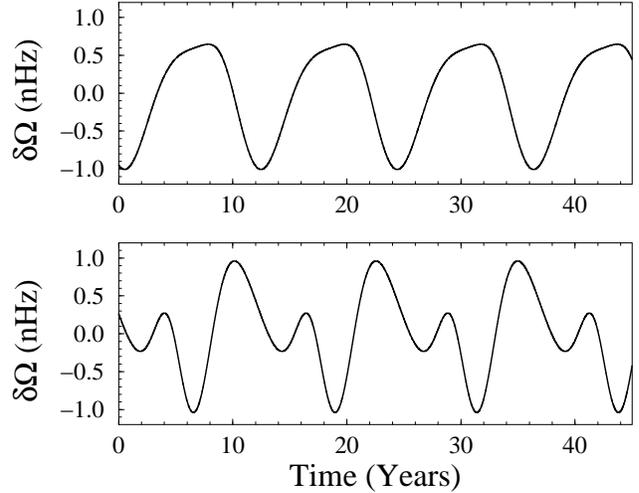}}
\caption{\label{time.series}
`Period halving' at $r=0.68$ and latitude $30^{\circ}$.
The panels correspond, from top to bottom, to $R_{\alpha} = -6.0$ and
$-11.0$ respectively, and display
increasing relative amplitudes of the secondary oscillations.
The remaining parameter values
are as in Fig.\ {\protect
\ref{dp_r=0.64_Calph=-11.0_iangalp=11_p=-1.0_Comega=44000_ialp=3_mesh=061x101_pr=1.0_ivac=2_eta=0_NS=1_rot=MDI2_rho=1.butterfly_bp}}.
}
\end{figure}

We also found the model to be capable of producing
spatiotemporal fragmentation
near the base of the convection zone,
i.e.\ we found there different dynamical modes of
behaviour in the differential rotation, including
oscillations with reduced periods, as well as
non-periodic variations as in CTM.
These coexist with the near-surface torsional oscillations of the form
shown in Figs.\
\ref{dp_r=0.64_Calph=-11.0_iangalp=11_p=-1.0_Comega=44000_ialp=3_mesh=061x101_pr=1.0_ivac=2_eta=0_NS=1_rot=MDI2_rho=1.velocity.R=0.95}
and
\ref{dp_r=0.64_Calph=-9.0_iangalp=11_p=-1.0_Comega=44000_ialp=3_mesh=061x101_pr=1.0_ivac=2_eta=0_NS=1_rot=GONG_rho=1.velocity.R=0.95}).
To demonstrate this, we have plotted in
Figs.\
\ref{dp_r=0.64_Calph=-11.0_iangalp=11_p=-1.0_Comega=44000_ialp=3_mesh=061x101_pr=1.0_ivac=2_eta=0_NS=1_rot=MDI2_rho=1.velocity_radial_latitude=30}
and
\ref{dp_r=0.64_Calph=-9.0_iangalp=11_p=-1.0_Comega=44000_ialp=3_mesh=061x101_pr=1.0_ivac=2_eta=0_NS=1_rot=GONG_rho=1.velocity_radial_latitude=30}
the radial contours of the angular velocity residuals $\delta \Omega$
as a function of time for a cut at latitude $30^{\circ}$.
To demonstrate the period halving produced by the fragmentation more clearly,
we have also plotted in
Fig.\ \ref{time.series}
the residuals $\delta \Omega$ at different values
of $R_\alpha$.

\begin{figure}[!htb]
\centerline{\def\epsfsize#1#2{0.38#1}\epsffile{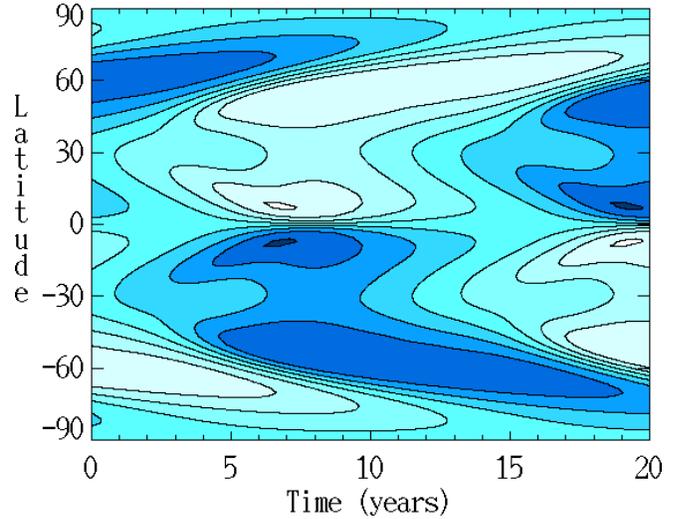}}
\caption{\label{dp_r=0.64_Calph=-11.0_iangalp=11_p=-1.0_Comega=44000_ialp=3_mesh=061x101_pr=1.0_ivac=2_eta=0_NS=1_rot=MDI2_rho=1.butterfly_bp_bottom}
Butterfly diagram of the toroidal component of the
magnetic field $\vec{B}$ at fractional radius $r=0.68$
for the MDI data, showing no signs of fragmentation.
Parameter values
are as in Fig.\ {\protect
\ref{dp_r=0.64_Calph=-11.0_iangalp=11_p=-1.0_Comega=44000_ialp=3_mesh=061x101_pr=1.0_ivac=2_eta=0_NS=1_rot=MDI2_rho=1.butterfly_bp}}.
}
\end{figure}

\begin{figure}[!htb]
\centerline{\def\epsfsize#1#2{0.38#1}\epsffile{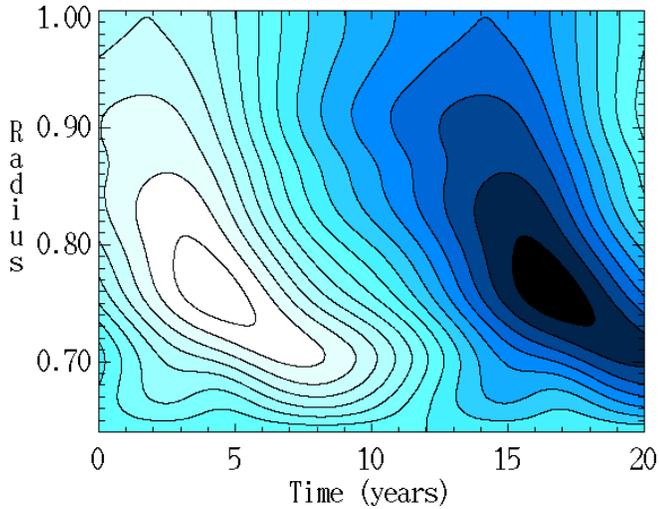}}
\caption{\label{dp_r=0.64_Calph=-11.0_iangalp=11_p=-1.0_Comega=44000_ialp=3_mesh=061x101_pr=1.0_ivac=2_eta=0_NS=1_rot=MDI2_rho=1.butterfly_bp_radial}
Radial contours of the toroidal component of the
magnetic field $\vec{B}$ as a function of time for
a cut at latitude $30^{\circ}$,
for the MDI data, showing no signs of fragmentation.
This is the counterpart of the Fig.\ {\protect
\ref{dp_r=0.64_Calph=-9.0_iangalp=11_p=-1.0_Comega=44000_ialp=3_mesh=061x101_pr=1.0_ivac=2_eta=0_NS=1_rot=GONG_rho=1.velocity_radial_latitude=30}}
and clearly shows that the fragmentation occurs only for $\delta\Omega(r,\theta,t)$ and that
the
magnetic field retains its typical 22 year cycle.
Parameter values
are as in Fig.\ {\protect
\ref{dp_r=0.64_Calph=-11.0_iangalp=11_p=-1.0_Comega=44000_ialp=3_mesh=061x101_pr=1.0_ivac=2_eta=0_NS=1_rot=MDI2_rho=1.butterfly_bp}}.
}
\end{figure}

We also find that, in all cases, torsional oscillations with the same period persist
in all the regions above the fragmentation level
as can, for example, be seen from Figs.\
\ref{dp_r=0.64_Calph=-11.0_iangalp=11_p=-1.0_Comega=44000_ialp=3_mesh=061x101_pr=1.0_ivac=2_eta=0_NS=1_rot=MDI2_rho=1.velocity_radial_latitude=30}
and
\ref{dp_r=0.64_Calph=-9.0_iangalp=11_p=-1.0_Comega=44000_ialp=3_mesh=061x101_pr=1.0_ivac=2_eta=0_NS=1_rot=GONG_rho=1.velocity_radial_latitude=30}.
The butterfly diagrams for the toroidal magnetic field, on the other
hand, keep the same period and qualitative form throughout the
convection zone, including the spatial regions that experience
fragmentation, as shown in
Figs.\
\ref{dp_r=0.64_Calph=-11.0_iangalp=11_p=-1.0_Comega=44000_ialp=3_mesh=061x101_pr=1.0_ivac=2_eta=0_NS=1_rot=MDI2_rho=1.butterfly_bp_bottom}
and
\ref{dp_r=0.64_Calph=-11.0_iangalp=11_p=-1.0_Comega=44000_ialp=3_mesh=061x101_pr=1.0_ivac=2_eta=0_NS=1_rot=MDI2_rho=1.butterfly_bp_radial}.
Thus the fragmentation in the angular velocity residuals do not seem
to be present in the magnetic field.  Fig.\
\ref{dp_r=0.64_Calph=-11.0_iangalp=11_p=-1.0_Comega=44000_ialp=3_mesh=061x101_pr=1.0_ivac=2_eta=0_NS=1_rot=MDI2_rho=1.butterfly_bp_radial}
shows clearly that the fragmentation only occurs
in $\delta\Omega$ and
that the magnetic field retains its typical 22 year cycle
throughout the dynamo region.

\begin{figure}[!htb]
\centerline{\def\epsfsize#1#2{0.42#1}\epsffile{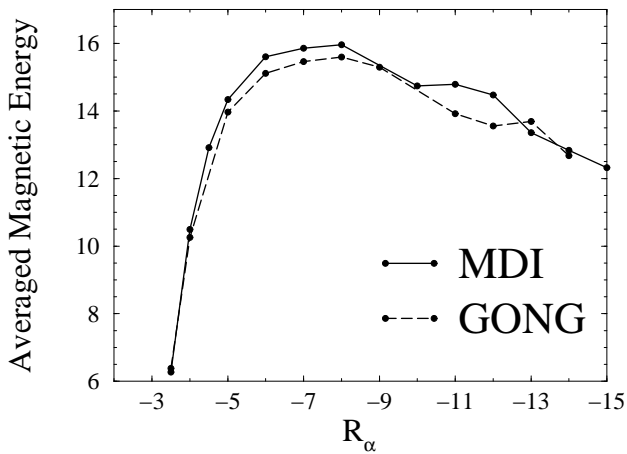}}
\caption{\label{magnetic.energy.averaged.Calpha.GONG-MDI}
The average magnetic energy $E_M$ as
a function of $R_\alpha$, for both MDI and the
GONG data.
The parameter values are
$P_r=1.0$ and ${R_{\omega}}=44000$.
}
\end{figure}

\begin{figure}[!htb]
\centerline{\def\epsfsize#1#2{0.45#1}\epsffile{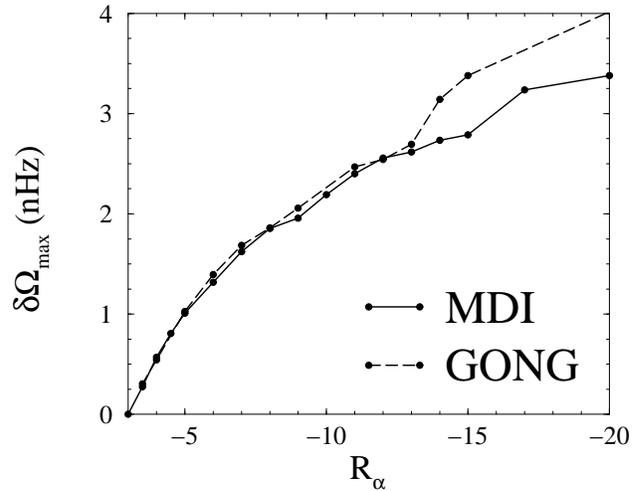}}
\caption{\label{maxima.amplitude.torsional.oscillations.30.degrees.GONG-MDI}
Comparison of the maximum of $\delta\Omega(r=0.95 R_{\odot},\theta)$ for both the MDI and
the GONG data sets.
The parameter values are
$P_r=1.0$ and ${R_{\omega}}=44000$.
}
\end{figure}

We also  made a detailed study of the magnetic field
evolution and the magnitude of torsional oscillations in each case.
Fig.\ \ref{magnetic.energy.averaged.Calpha.GONG-MDI} summarises the
results of the average magnetic energy
as a function of $R_\alpha$, for both the MDI and the
GONG data, and demonstrates that  both data sets produce almost the same
average magnetic energy for a given
$R_\alpha$.
Similarly, we have plotted in Fig.\
\ref{maxima.amplitude.torsional.oscillations.30.degrees.GONG-MDI},
the maxima (of the absolute value) of the residuals of the differential rotation,
$\delta\Omega$, as a function of $R_\alpha$, for both MDI and the
GONG data, which shows that for large values of $| R_\alpha |$,
the GONG data produce larger residuals.

We end this section by summarising the qualitative modes of
spatiotemporal behaviour our detailed numerical
results have produced,
for both MDI and GONG data.
Our results show that our model is capable of producing
three qualitatively different spatiotemporal modes of behaviour:
(i) regimes in which there is no
deformation of the
torsional oscillation bands in the ($r,t$) plane, and hence
no changes in phase or period through
the convection zone;
(ii) regimes in which there is deformation in the
oscillatory bands in the ($r,t$) plane,
resulting in changes in the phase
of the oscillations, but no changes in their period
and (iii) regimes with spatiotemporal fragmentation,
resulting in changes both in phase as well as in period/behaviour of
oscillations.
An example of the difference between such regions
is given in Fig.\
\ref{dp_r=0.64_Calph=-11.0_iangalp=11_p=-1.0_Comega=44000_ialp=3_mesh=061x101_pr=1.0_ivac=2_eta=0_NS=1_rot=MDI2_rho=1.velocity_radial_latitude=30},
which shows a spatiotemporal fragmentation resulting in period halving,
as well as a phase change between the oscillations near the surface and those
deeper down in the convection zone.
We should emphasise that by {\it period} we mean here the time between
the maxima (minima), rather than time between repeated
sequences. We note that given the presence of noise,
the error bars induced by the inversion
and the shortness of the observational interval,
the time between
the maxima (minima) is
the most relevant quantity from the point of view of
comparison with observations.

\begin{figure}[!htb]
\centerline{\def\epsfsize#1#2{0.47#1}\epsffile{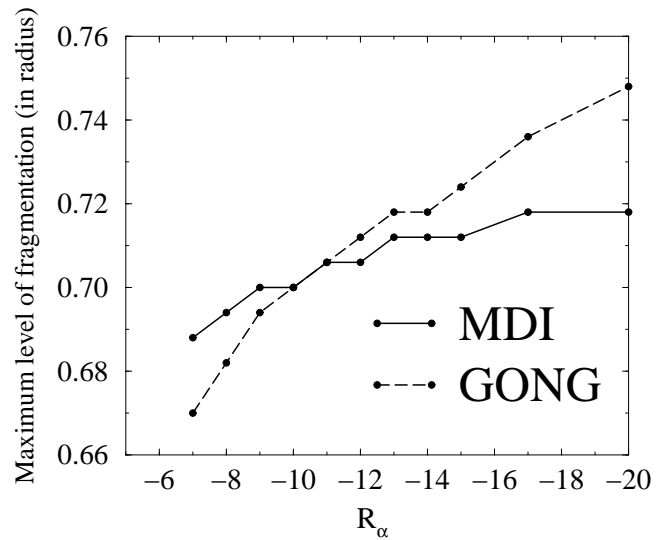}}
\caption{\label{fragmentation.level.MDI.GONG.sin4}
Variation of fragmentation level (the maximum radius for which more than
one fundamental period exists) as a function of the control parameter
$R_\alpha$ for both the MDI and
the GONG data sets.
The parameter values are
$P_r=1.0$ and ${R_{\omega}}=44000$.
}
\end{figure}

\begin{figure}[!htb]
\centerline{\def\epsfsize#1#2{0.47#1}\epsffile{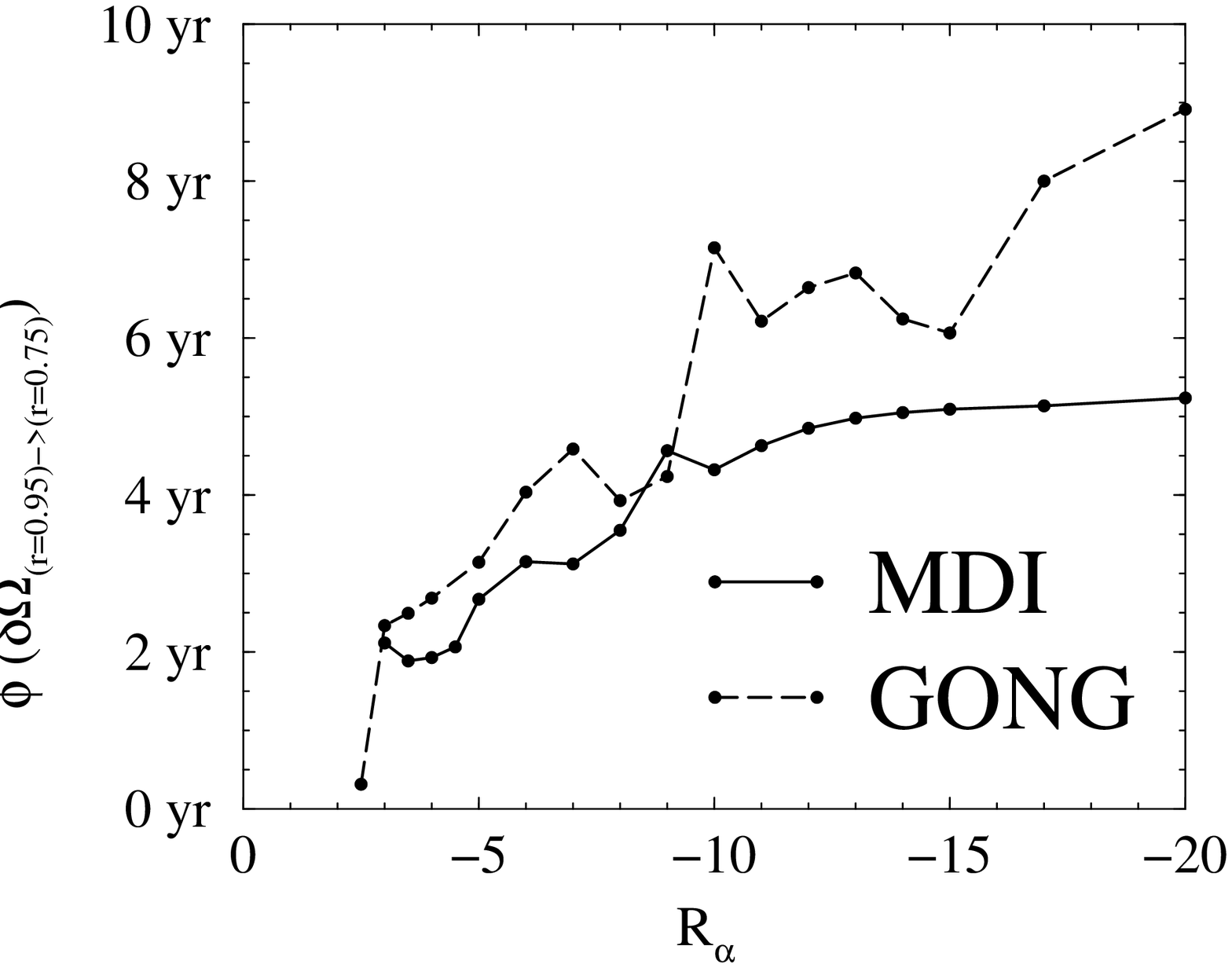}}
\caption{\label{phase.MDI.GONG.sin4}
Variation of phase shift $\phi$ as a function of the control parameter
$R_\alpha$ for both the MDI and
the GONG data sets. The phase shift evaluated by comparing the residual time series
$\delta\Omega(r=0.95 R_{\odot},\theta=30^{\circ})$ and $\delta\Omega(r=0.75
R_{\odot},\theta=30^{\circ})$.
A positive phase shift $\phi$ indicates a forward shift of the
torsional oscillation bands (top to bottom) and
a negative phase shift an opposite one.
The parameter values are
$P_r=1.0$ and ${R_{\omega}}=44000$.
}
\end{figure}

Our results also show that using the zero order rotation
produced by both the MDI and GONG data gives rise to
qualitatively similar results regarding both the spatiotemporal fragmentation
and the torsional oscillations. There are, however,
detailed quantitative
differences. An example of this relates to the case (ii) above, where the location of
where the fragmentation occurs is higher for MDI
than for GONG at higher values of $|R_\alpha|$
(see Fig.\ \ref{fragmentation.level.MDI.GONG.sin4}).
Furthermore, the phase shift is less  pronounced for the MDI
case than for the GONG case, specially at higher
$| R_\alpha| $ values (see Fig.\ \ref{phase.MDI.GONG.sin4}).

\section{Robustness of the model}

As noted above, the model of the solar dynamo used by CTM to demonstrate
the presence of torsional oscillations and
spatiotemporal fragmentation is inevitably
approximate in nature and contains major simplifications and parameterizations.
It is therefore important to ask to what
extent the spatiotemporal fragmentation, as well
as the qualitative features of torsional oscillations
found in CTM, depend on the details of
the model employed.

In order to go some way towards answering these questions, we
shall in this section study the robustness of this
model with respect to plausible changes to its main
ingredients, considering
these changes one by one.

A feature of the model considered by CTM is the
presence of radial variations in both $\alpha$ and $\eta$, and in
particular the fact that $\alpha$ was taken to be zero in $r\leq 0.70$.
An obvious question is whether this variation is the source of the
fragmentation described above, for example does it cause  the dynamo region
essentially to
consist of two disparate parts, each with its own set of properties?
This is perhaps unlikely, as there is no independent dynamo action in
$r\leq 0.7$, as $\alpha=0$ there. Nevertheless, to resolve this issue,
we study the effects of changes in the radial variations of both
$\alpha$ and $\eta$ and subsequently examine a model in which neither
vary radially.

Given the qualitative similarity of the results produced
using the MDI and the GONG data,
and in order to keep the extent of the computations
within tractable bounds, we consider only
the MDI data in the following sections.
\subsection{Robustness with respect to changes in the $\alpha$ profile}
\begin{figure}[!htb]
\centerline{\def\epsfsize#1#2{0.38#1}\epsffile{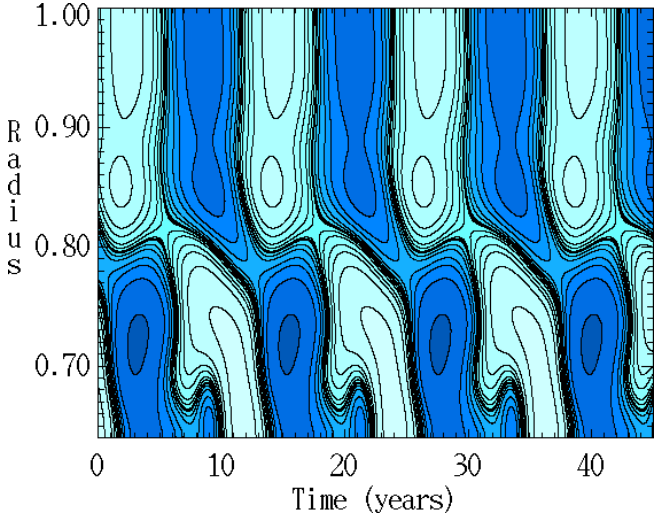}}
\caption{\label{dp_r=0.64_Calph=-11.0_iangalp=11_p=-1.0_Comega=44000_ialp=0_mesh=061x101_pr=1.0_ivac=2_eta=0_NS=1_rot=MDI2_rho=1.velocity_radial_latitude=30}
Radial contours of the angular velocity residuals $\delta \Omega$
as a function of time for a cut at latitude $30^{\circ}$.
The parameter values
are $R_\omega=44000$, $R_\alpha=-11.0$ and $Pr=1.0$
with  $\alpha_r(r)=1$.
}
\end{figure}

We studied this by
setting $\alpha_r=1$ throughout the computational region, which changes the
$\alpha$ profile substantially.
In spite of this, we found qualitatively similar forms of butterfly diagrams,
torsional oscillations and
spatiotemporal fragmentation in this case as in the case of CTM.
An example of such fragmentation
in the radial contours of $\delta \Omega$ is
given in Fig.\
\ref{dp_r=0.64_Calph=-11.0_iangalp=11_p=-1.0_Comega=44000_ialp=0_mesh=061x101_pr=1.0_ivac=2_eta=0_NS=1_rot=MDI2_rho=1.velocity_radial_latitude=30}.

We note that the change in the $r$-dependence of $\alpha$ does not change
the calibration of $R_\omega$ required to produce a cycle period of 22 years.
It does, however, seem to change the
details of the surface torsional oscillations,
making them somewhat less realistic. It also, importantly from the point of
view of observations, enhances the phase shift between the torsional oscillations
at the top and below the deformation, to a phase change of $\phi= -\pi$ as can
be seen from Fig.\ \ref{phase.others1}.
Note, however,
that the phase shift $\phi$ changes continuously with $R_{\alpha}$.

\subsection{Robustness with respect to $\eta$ profile}
To obtain some idea of the sensitivity of the model to the chosen
radial dependence of the turbulent diffusion coefficient $\eta$,
we then examined a model in which $\tilde\eta (r)=1$ throughout. In this way
we removed from the model the inhomogeneity associated with $\eta$ at the
base of the convection zone.

We found that with this form of $\eta$,
$R_\omega$ needs to be recalibrated
to $R_\omega =60000$ in order
to produce toroidal field butterfly diagrams at the surface with a period of 22 years.

\begin{figure}[!htb]
\centerline{\def\epsfsize#1#2{0.38#1}\epsffile{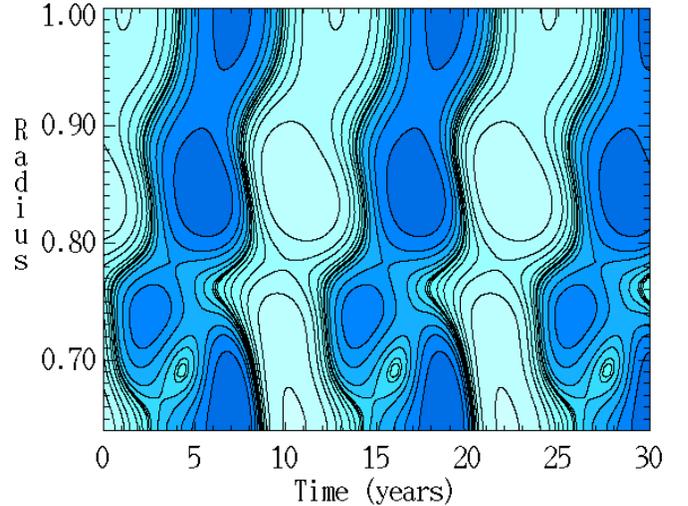}}
\caption{\label{dp_r=0.64_Calph=-16.0_iangalp=11_p=-1.0_Comega=60000_ialp=3_mesh=061x101_pr=1.0_ivac=2_eta=3_NS=1_rot=MDI2_rho=1.velocity_radial_latitude=30}
Radial contours of the angular velocity residuals $\delta \Omega$
as a function of time for a cut at latitude $30^{\circ}$.
The parameter values
are $R_\omega=60000$, $R_\alpha=-16.0$ and $Pr=1.0$
with $\tilde\eta(r)=1$.
}
\end{figure}

With this value of $R_\omega$, we again found qualitatively similar forms of
spatiotemporal fragmentation
in this case, to those obtained by CTM.
An example is given in
Fig.\
\ref{dp_r=0.64_Calph=-16.0_iangalp=11_p=-1.0_Comega=60000_ialp=3_mesh=061x101_pr=1.0_ivac=2_eta=3_NS=1_rot=MDI2_rho=1.velocity_radial_latitude=30}.

We note that, compared with
the behaviour of the original
model depicted in
Figs.\
\ref{dp_r=0.64_Calph=-9.0_iangalp=11_p=-1.0_Comega=44000_ialp=3_mesh=061x101_pr=1.0_ivac=2_eta=0_NS=1_rot=GONG_rho=1.butterfly_bp},
this model produces more realistic torsional
oscillations at the surface than the model
with the radial dependence of $\alpha$ removed.
However, the butterfly diagrams of the toroidal
magnetic field near the surface look slightly less realistic
than those for the original model shown in
Fig.\
\ref{dp_r=0.64_Calph=-11.0_iangalp=11_p=-1.0_Comega=44000_ialp=3_mesh=061x101_pr=1.0_ivac=2_eta=0_NS=1_rot=MDI2_rho=1.butterfly_bp}.
But, it must be borne in mind that in the absence of a definitive sunspot
model, it is not obvious at which depth the toroidal field patterns
correspond to the observed butterfly diagram.
Furthermore, as a result of the removal of the radial variation of $\eta$ in this model,
the effective magnetic turbulent diffusivity
changes and consequently the range of
$R_\alpha$ values for spatiotemporal fragmentation are
somewhat higher than for the model
employed by CTM. We note also that the fragmentation is now present
at higher radius and is in this case detached from the lower boundary.

\subsection{Robustness to simultaneous removal of
radial dependence of $\alpha$ and $\eta$}

\begin{figure}[!htb]
\centerline{\def\epsfsize#1#2{0.38#1}\epsffile{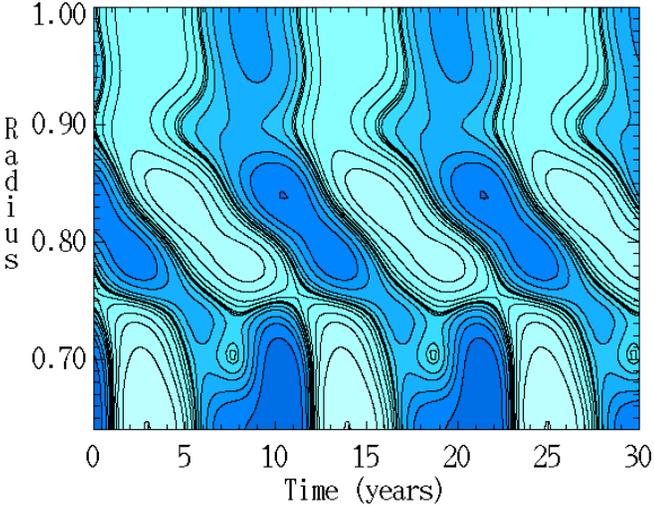}}
\caption{\label{dp_r=0.64_Calph=-15.0_iangalp=11_p=-1.0_Comega=60000_ialp=0_mesh=061x101_pr=1.0_ivac=2_eta=3_NS=1_rot=MDI2_rho=1.velocity_radial_latitude=30}
Radial contours of the angular velocity residuals $\delta \Omega$
as a function of time for a cut at latitude $30^{\circ}$.
Parameter values
are $R_\omega=60000$, $R_\alpha=-15.0$ and $Pr=1.0$
with $\eta(r,\theta)=1$ and $\alpha_r(r)=1$.
}
\end{figure}

In the last two sections we studied the robustness of
the spatiotemporal fragmentation and the torsional oscillations
with respect to the removal  of the radial dependence
of $\alpha$ and $\eta$ in turn.
Interestingly, in both cases we found
the spatiotemporal fragmentation to persist.
Here we investigate a model in which both $\alpha$ and $\eta$
are independent of radius, so that (unrealistically) the convective
region and overshoot layer are only distinguished by the nature of the
given zero order rotation law.
Thus we considered a modification of the model
considered by CTM, given by taking $\alpha_r=\tilde\eta(r)=1$.
Perhaps surprisingly, we still found
spatiotemporal fragmentation, as can be seen
in Fig.\
\ref{dp_r=0.64_Calph=-15.0_iangalp=11_p=-1.0_Comega=60000_ialp=0_mesh=061x101_pr=1.0_ivac=2_eta=3_NS=1_rot=MDI2_rho=1.velocity_radial_latitude=30}.

We note, however, that
in this case
both the butterfly diagrams
of the toroidal field and the
torsional oscillations near the surface
look less realistic than the original model
of Sect.\ \ref{tos}.
Our point is however that spatiotemporal fragmentation appears to be a
general property of the type of dynamo models studied, and is not
crucially dependent
on imposed spatial structures in the coefficients $\alpha$ and $\eta$,
that might be thought to distinguish the lower part of the dynamo
region from the subsurface layers. We also  note that again the fragmentation region
is detached from the lower boundary (suggesting that this is a feature related
to the $\eta$ profile) and extends to an even greater height, $r=0.71 R_{\odot}$.
It is interesting that, within our
limited experimentation,
the details of the fragmentation, but not its existence,
seem to be quite sensitive to the $\eta$ profile.

\begin{figure}[!htb]
\centerline{\def\epsfsize#1#2{0.42#1}\epsffile{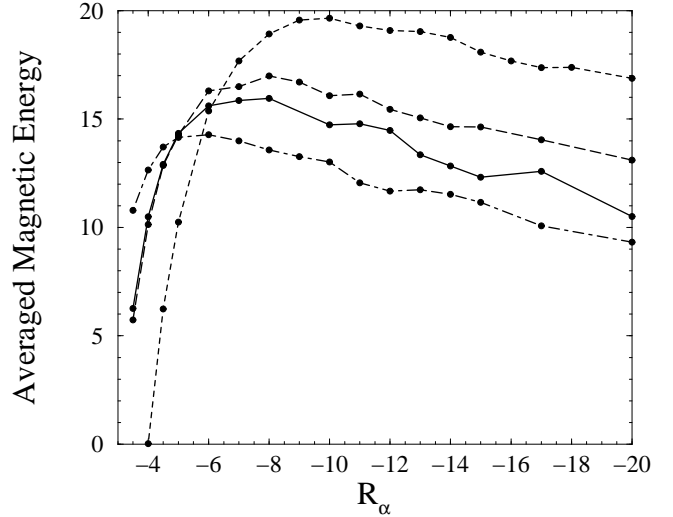}}
\caption{\label{magnetic.energy.averaged.Calpha.other1}
Comparison of the average magnetic energy $E_M$
as a function of $R_\alpha$ for models with
different $\alpha$ profiles.
The parameter values are $P_r=1.0$ and ${R_{\omega}}=44000$.
The continuous curve represents the results of the model
with $\alpha_r = f(r), \eta (r,\theta) = \eta (r)$,
the long-dashed curve the model with $\alpha_r = 1 = \eta (r,\theta)$,
the short-dashed curve the model with $\eta (r,\theta) = 1$ and
the dotted-dashed curve the model with
$\alpha_r = 1$.
}
\end{figure}

\begin{figure}[!htb]
\centerline{\def\epsfsize#1#2{0.45#1}\epsffile{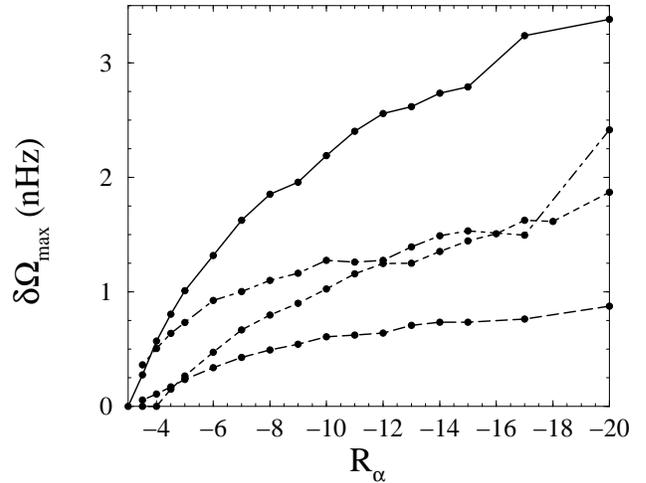}}
\caption{\label{maxima.amplitude.torsional.oscillations.30.degrees.others1}
Comparison of the maximum of $\delta\Omega(r=0.95 R_{\odot},\theta)$ for  models with
different $\alpha$ profiles.
The parameter values and the characterisation of the curves
is as in  {\protect Fig.\ \ref {magnetic.energy.averaged.Calpha.other1}}.
}
\end{figure}

We also studied the average magnetic energy and
the residuals,
$\delta\Omega$, as a function of $R_\alpha$
with different $\alpha$ and $\eta$ profiles
and these are depicted in Figs.\ \ref{magnetic.energy.averaged.Calpha.other1} and
\ref{maxima.amplitude.torsional.oscillations.30.degrees.others1}.
As can be seen, the values of $\delta \Omega$ are larger  for the
case with radial variations of both $\alpha$ and $\eta$ present,
whereas the average magnetic energy is largest for the
case with uniform  $\eta$.

\subsection{Robustness with respect to $\alpha$ quenching}
Even though the model used by CTM was nonlinear (via
the Lorentz force and the subsequent $v'$ term in
the induction equation), the
magnitude of $\alpha$ was fixed {\it ab initio}. We note that
the form of
$\alpha_r$ in the original model was initially chosen to
represent implicit  strong $\alpha$-quenching
occurring in the overshoot layer, in that $\alpha=0$ in $r\leq 0.7$.
In this section we
study the effects of having an additional nonlinearity in the form of an
$\alpha$ quenching given by
\begin{equation}
\label{quenching}
\tilde\alpha= \alpha_r(r) \frac{\sin^4\theta\cos\theta}{1+ g|{\bf B}^2|}
\end{equation}
where $g$ is a quenching factor, in addition
to the nonlinearity given by Eq.\ (\ref{NS}).

In this connection it may be noted that there is an ongoing controversy
regarding the nature and strength of $\alpha$-quenching.
This is not the place to rehearse the arguments for and against `strong'
alpha-quenching; the issue is unresolved, and we use (\ref{quenching})
as a commonly adopted nonlinearity.

We found that for values of $g \lta 0.1$ (depending somewhat on $R_\alpha$ and $P_r$),
the model continues to produce torsional
oscillations and spatiotemporal fragmentation.
\subsection{Robustness with respect to the Prandtl number $P_r$}

In the results obtained by CTM,
the Prandtl number was taken as $P_r=1$.
In this section we study the effects of changing $P_r$
on the nature of spatiotemporal fragmentation and
torsional oscillations.

Our studies show that, for our model,
both spatiotemporal fragmentation
and torsional oscillations
persist for values of Prandtl number given by
$P_r \gta 0.4$ (depending on $R_\alpha$). Around $P_r\approx 0.4$, however, a sudden transition
seems to occur, such that below this value spatiotemporal fragmentation
as well as
coherent surface torsional oscillations
seem to be absent.

\begin{figure}[!htb]
\centerline{\def\epsfsize#1#2{0.45#1}\epsffile{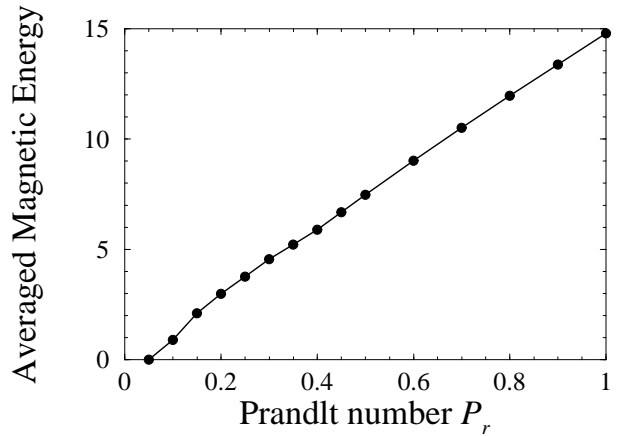}}
\caption{\label{magnetic.energy.averaged.prandlt.MDI.sin4}
The average magnetic energy $E_M$ as
a function of $P_r$.
The parameter values
are $R_\omega=44000$, $R_\alpha=-15.0$ and $Pr=0.5$.
}
\end{figure}

\begin{figure}[!htb]
\centerline{\def\epsfsize#1#2{0.45#1}\epsffile{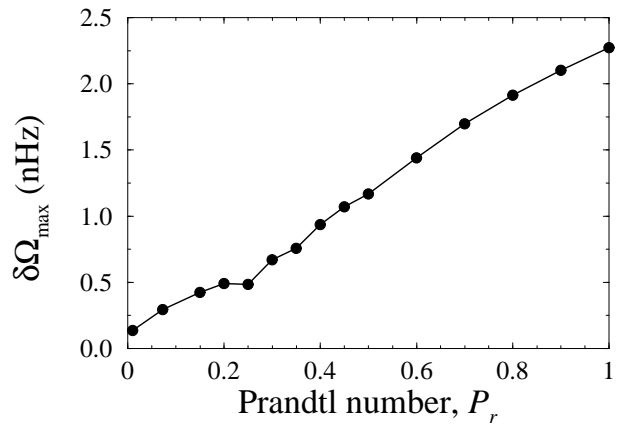}}
\caption{\label{maxima.amplitude.torsional.oscillations.30.degrees.function.prandlt}
The maximum of $\delta\Omega(r=0.95 R_{\odot},\theta)$ as
a function of $P_r$ when $R_\omega=44000$, $R_\alpha=-15.0$.}
\end{figure}

We also made a detailed study of the magnetic field
strength as well as of the magnitude of the torsional oscillations,
for different values of
Prandtl number $P_r$. Figs.\ \ref{magnetic.energy.averaged.prandlt.MDI.sin4} and
\ref{maxima.amplitude.torsional.oscillations.30.degrees.function.prandlt}
show the behaviours
of the average magnetic energy and the
residuals of the differential rotation,
$\delta\Omega$, for fixed $R_\alpha$,
for different values of
Prandtl number $P_r$.
This is in agreement with the results of K\"uker et al.\ 1996,
who studied solar torsional oscillations in a model with
magnetic quenching of the Reynolds stresses  (the `$\Lambda$-effect'),
with both studies showing a linear growth of the amplitude of
the torsional oscillations as a function of (small)
Prandtl number. In their case the amplitudes of the torsional oscillations
for $P_r \lta 1$ are weaker than observed, whereas in our model
they are within physically more reasonable ranges, $\delta \Omega \approx 0.5$--$1$ nHz.
However, the restriction to uniform density makes it uncertain how important
these amplitude differences really are.

\begin{figure}[!htb]
\centerline{\def\epsfsize#1#2{0.42#1}\epsffile{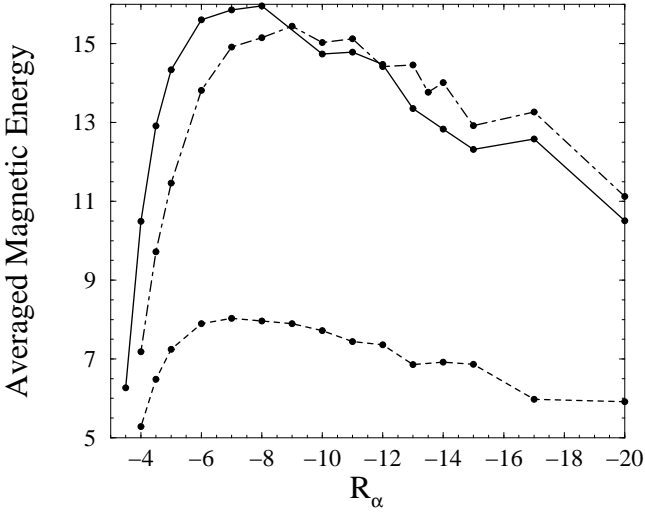}}
\caption{\label{magnetic.energy.averaged.Calpha.other2}
Comparison of the average magnetic energy $E_M$ for models with
$\alpha$--quenching and smaller $P_r$.
In this figure, ${R_{\omega}}=44000$,
the continuous curve corresponds to the model with
$\alpha_r = f(r), \eta (r,\theta) = \eta (r), P_r =1.0$,
the dot--dashed line is for the model with
$\alpha$--quenching and the dashed curve for the model with
$P_r =0.5$ and no $\alpha$-quenching.
}
\end{figure}

\begin{figure}[!htb]
\centerline{\def\epsfsize#1#2{0.45#1}\epsffile{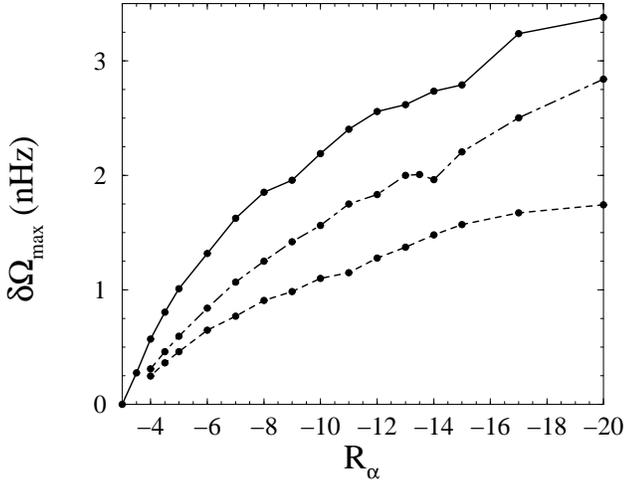}}
\caption{\label{maxima.amplitude.torsional.oscillations.30.degrees.others2}
Comparison of the maximum of $\delta\Omega(r=0.95 R_{\odot},\theta)$ for
models with
$\alpha$--quenching and smaller values of $P_r$.
In this figure, ${R_{\omega}}=44000$,
the continuous curve corresponds the model with
$\alpha_r = f(r), \eta (r,\theta) = \eta (r)$,
$P_r =1.0$,
the dot--dashed line represents the model with
$\alpha$--quenching and the dashed curve the model with
$P_r =0.5$.
}
\end{figure}

Finally we also studied the average magnetic energy and the
maximum value of the residuals,
$\delta\Omega$, as a function of $R_\alpha$
for a number of cases,
including a nonlinearly quenched $\alpha$,
as shown in Figs.\ \ref{magnetic.energy.averaged.Calpha.other2} and
\ref{maxima.amplitude.torsional.oscillations.30.degrees.others2}.
The decrease in $P_r$ results in the reduction of
the average magnetic energy and the residuals,
as can be seen from these figures.

\subsection{Variations in fragmentation level and phase}
Finally, in this section we briefly summarise our detailed results concerning the
variations in the spatiotemporal fragmentation
level (i.e. the maximum radius at which more than one
period exists) as well as changes in the phase $\phi$,
as a function of the control parameters
$R_\alpha$ and $P_r$ for models with
different $\alpha$ and $\eta$ profiles
as well as those with nonlinear $\alpha$--quenching.

\begin{figure}[!htb]
\centerline{\def\epsfsize#1#2{0.47#1}\epsffile{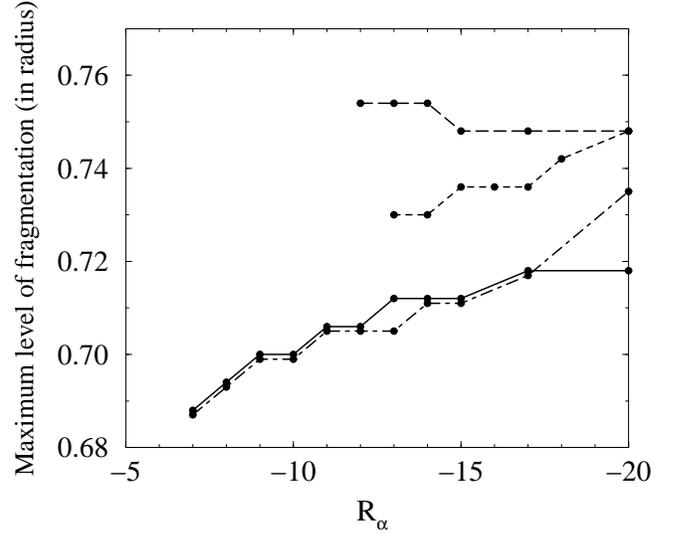}}
\caption{\label{fragmentation.others1}
Variation of spatiotemporal fragmentation level
for different $\alpha$ and $\eta$ profiles.
Here $\Omega(r,\theta)$ is given by the MDI data,
$P_r=1.0$, $R_{\omega}=44000$ and
the continuous curve gives the results for the model
with $\alpha_r = f(r), \eta (r,\theta) = \eta (r)$,
the long-dashed curve the model with $\alpha_r =\eta (r,\theta) = 1$,
the short-dashed curve the model with $\eta (r,\theta) = 1$ and
the dotted-dashed curve the model with
$\alpha_r = 1$ (slightly displaced downwards for clarity).
}
\end{figure}

\begin{figure}[!htb]
\centerline{\def\epsfsize#1#2{0.47#1}\epsffile{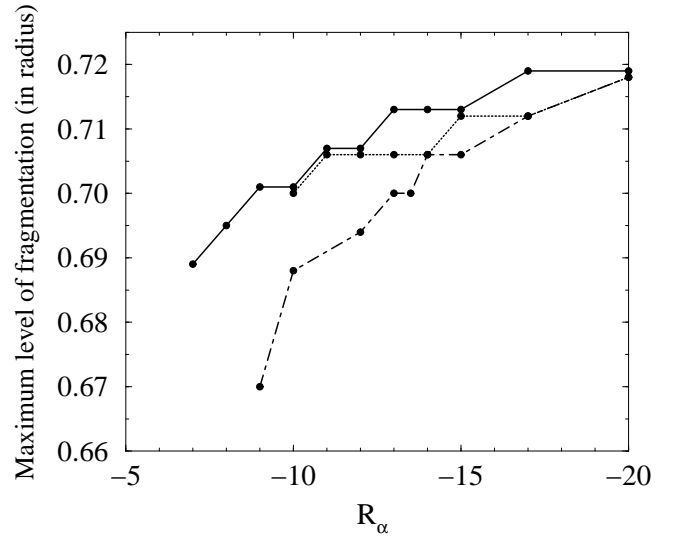}}
\caption{\label{fragmentation.others2}
Variation of fragmentation level as a function
of $R_\alpha$, in the  presence of $\alpha$--quenching
as  $P_r$ is reduced.
Here ${R_{\omega}}=44000$ and
the continuous curve gives the results for the model with
$\alpha_r = f(r), \eta (r,\theta) = \eta (r), P_r =1.0$
(slightly displaced upwards for clarity),
the dot--dashed line represents the model with
$\alpha$--quenching and the dashed curve the model with
$P_r =0.5$.
}
\end{figure}

Fig.\ \ref{fragmentation.others1} shows the
radius of the upper boundary of the region of
spatiotemporal fragmentation
as a function of
$R_\alpha$ for different $\alpha$ and $\eta$ profiles
at $P_r =1$. As can be seen, for intermediate
values of $|R_\alpha|$, the models with
uniform $\eta$
have a higher fragmentation level. Also they
also have fragmentation at values of $|R_\alpha|$
significantly larger than for the model with a radial
dependence of $\eta$. The higher upper boundary of the
fragmentation region for such models
is associated with the position of the fragmented cells,
which are now detached from the bottom boundary, unlike the
basic models described earlier (cf.\ Figs.\
\ref{dp_r=0.64_Calph=-11.0_iangalp=11_p=-1.0_Comega=44000_ialp=3_mesh=061x101_pr=1.0_ivac=2_eta=0_NS=1_rot=MDI2_rho=1.velocity_radial_latitude=30},
\ref{dp_r=0.64_Calph=-16.0_iangalp=11_p=-1.0_Comega=60000_ialp=3_mesh=061x101_pr=1.0_ivac=2_eta=3_NS=1_rot=MDI2_rho=1.velocity_radial_latitude=30}
and
\ref{dp_r=0.64_Calph=-15.0_iangalp=11_p=-1.0_Comega=60000_ialp=0_mesh=061x101_pr=1.0_ivac=2_eta=3_NS=1_rot=MDI2_rho=1.velocity_radial_latitude=30}).
At small
values of $|R_\alpha|$ the removal of the radial
dependence on the $\alpha$ profile does not change
the maximum height at which fragmentation occurs, and even at higher
$|R_\alpha|$ it does so only slightly.

We have also studied the position of the boundary of the
fragmentation region
in models with $\alpha$--quenching as well as
lower values of  $P_r$ and the
results are shown in
Fig.\ \ref{fragmentation.others2}.
It can clearly be seen that the quenching of $\alpha$
changes significantly the position and amplitude of
the fragmentation. Not only does it start at higher
$|R_\alpha|$ but it also  extends closer to the bottom boundary.
The dependence on the Prandtl number is somehow
less pronounced. Fragmentation again sets in
starts at higher values of $|R_\alpha|$
but the position of the fragmentation region is almost the same
as that of the basic model with $P_r=1$.

\begin{figure}[!htb]
\centerline{\def\epsfsize#1#2{0.47#1}\epsffile{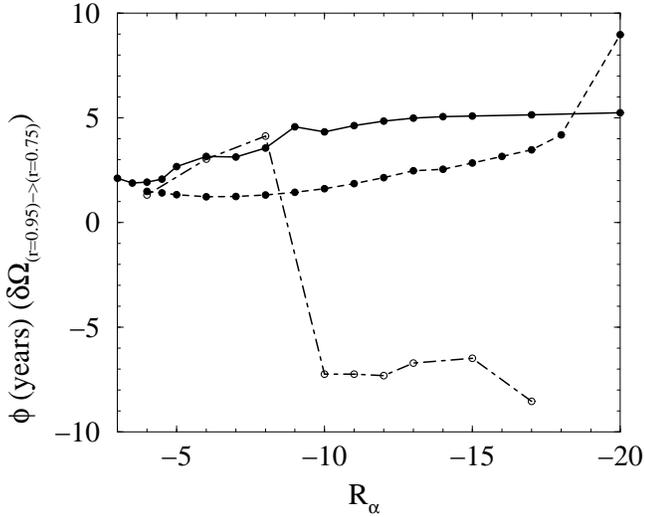}}
\caption{\label{phase.others1}
Variation of the phase shift $\phi$ as a function of
$R_\alpha$ for different $\alpha$ and $\eta$ profiles.
Here $\Omega(r,\theta)$ is given by the MDI data,
$P_r=1.0$, $R_{\omega}=44000$ and
the continuous curve gives the results for the model
with $\alpha_r = f(r), \eta (r,\theta) = \eta (r)$,
the short-dashed curve the model with $\eta (r,\theta) = 1$ and
the dotted-dashed curve the model with
$\alpha_r = 1$. We were unable to calculate uniquely the
phase shift for the case without radial dependence in both
the $\alpha$ and $\eta$ profiles and so we have omitted it.
}
\end{figure}

Further, we studied the shift in phase, $\phi$, between the oscillations
near the top and bottom of the dynamo region.
Fig.\ \ref{phase.others1} shows the
phase shift $\phi$ as a function of
$R_\alpha$, for different $\alpha$ and $\eta$ profiles.
As can be seen, the phase shift can be negative or
positive. For the models with $\alpha_r = 1$, $\phi$ can be either
positive or negative, while for the other models
it is always positive. We note that the phases shifts
for small enough $|R_\alpha|$ are correspondingly small, that is the torsional oscillations
are all in phase throughout the convection zone.

\begin{figure}[!htb]
\centerline{\def\epsfsize#1#2{0.47#1}\epsffile{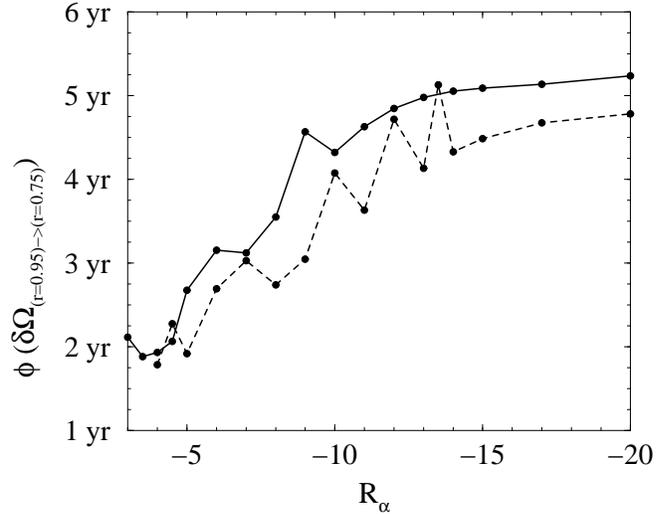}}
\caption{\label{phase.others2}
Variation of phase shift $\phi$ as a function of $R_\alpha$
in the presence of $\alpha$--quenching.
Here ${R_{\omega}}=44000$, the solid line gives the results for the model
with $\alpha$ quenching and the dashed line those for the model
with $\alpha$--quenching with $g=10^{-2}$.
}
\end{figure}

In Fig.~\ref{phase.others2} we
have plotted the
phase shifts for the $\alpha$--quenched
models. This shows that $\alpha$--quenching does
not dramatically change the behaviour (and the sign) of the phase
shifts (apart from reducing it somewhat), for nearly all values of
$R_\alpha$.

\section{Discussion}
We begin by acknowledging the considerable uncertainties and simplifications
which are associated with all solar dynamo models, including our own. In particular, in
the present context our assumption of uniform density may be important.

We have made a detailed study of a recent proposal
in which the recent results of
helioseismic observations regarding the
dynamical modes of behaviour in the
solar convection zone are accounted for in terms
of spatiotemporal fragmentation.
Originally, support for this scenario came from the
study of a particular two dimensional axisymmetric mean field
dynamo model operating in a spherical shell, with
a semi--open outer boundary condition,
which inevitably involved
a number of simplifying and somewhat arbitrary assumptions.
To demonstrate, to a limited extent at least,
the independence of our proposed mechanism
from the details of our model, we have shown that
this scenario is robust with respect to a number of
of plausible changes to the main ingredients of
the model, including the $\alpha$ and $\eta$ profiles as well as
the inclusion of nonlinear quenching and changes in the Prandtl number.
We have also shown the persistence of spatiotemporal fragmentation
with respect to the zero order angular velocity (by
considering both the MDI and the GONG data sets), as well
as under changes in the form of the factor $f(\theta)$ that
prescribes the latitudinal dependence of $\alpha$
(by considering $f(\theta) = \sin^4 \theta \cos \theta$).
We further found our model to be
capable of producing butterfly
diagrams which are in qualitative agreement
with the observations.
In this way we have found evidence
that spatiotemporal fragmentation
is not confined to our original model
only and that it can occur in
more general dynamo models.

Concerning our model, we should note that all our calculations
were done with semi--open outer boundary conditions
on the toroidal magnetic field (i.e. $\partial B/\partial r=0$).
In spite of relatively extensive searches,
we have so far been unable to
find spatiotemporal fragmentation
with pure vacuum boundary conditions ($B=0$).
It is interesting to note in this connection
that some examples of spatiotemporal
bifurcations in the literature (albeit in coupled maps)
also have open boundary conditions (see e.g.\ Willeboordse \& Kaneko 1994, Frankel et al.\
1994).
This may therefore be taken as some tentative support for the idea that
spatiotemporal fragmentation may require
some sort of open boundary conditions, or that at least it is easier to
occur in such settings.

We also made an extensive study of the spatial magnetic field
structure as well as the  nature of dynamical variations
in the differential rotation,
including amplitudes and phases, as a function of
depth and latitude.
Our results demonstrate the presence of
three main qualitative spatiotemporal regimes:
(i)
regimes where there is no deformation of the
torsional oscillations bands in the ($r,t$) plane, and hence
no changes in phase or period through
the convection zone; (ii)
regimes where there is spatiotemporal deformation in the
oscillatory bands in the ($r,t$) plane,
resulting in changes in the phase
of the oscillations, but no changes in their period,
and (iii) regimes with spatiotemporal fragmentation,
resulting in changes both in phase as well as period/behaviour of
oscillations, including
regimes that are markedly different from those
observed at the top, having either significantly reduced
periods or non-periodic modes of behaviour.
In all three cases, we found
that torsional oscillations, resembling those found near marginal
excitation, persisted
above the fragmentation level.

These modes of behaviour can in principle explain a
number of features that have been observed recently.
These include:

\begin{description}
\item[(a)]
Deformations of the type (ii) with or without
fragmentation of type (iii), can account for
the reversal of the phase of
the oscillatory behaviour above and below the tachocline,
as reported by Howe et al.\ (2000a).
\item[(b)]
Spatiotemporal fragmentation can
explain latitudinal dependence, as fragmentation occurs
in both radius/time as well as in latitude/time plots.
In particular, it can explain the possibility
of oppositely signed tachocline shear
at low and high latitudes, as has been
found in observations by Howe et al.\ (2000a).
To demonstrate this, we have plotted
in
Figs.\
\ref{dp_r=0.64_Calph=-11.0_iangalp=11_p=-1.0_Comega=44000_ialp=3_mesh=061x101_pr=1.0_ivac=2_eta=0_NS=1_rot=MDI2_rho=1.residual_snapshot}
and \ref{opposite.signed.tachocline.r=0.68.MDI.sin4.data} ,
the residuals of the differential rotation rate
with respect to the radius and the latitude.

\begin{figure}[!htb]
\centerline{\def\epsfsize#1#2{0.37#1}\epsffile{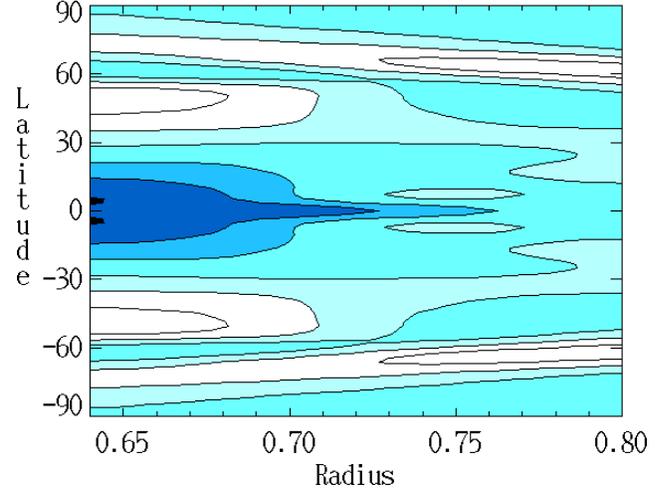}}
\caption{\label{dp_r=0.64_Calph=-11.0_iangalp=11_p=-1.0_Comega=44000_ialp=3_mesh=061x101_pr=1.0_ivac=2_eta=0_NS=1_rot=MDI2_rho=1.residual_snapshot}
Variation of the perturbation to the zero order rotation rate in latitude and time,
revealing the
migrating banded zonal flows, after the
transients have died out,
using the MDI data.
Observe that residuals with different signs can occur
over small latitude or radii bands.
Parameter values
are as in Fig.\ {\protect
\ref{dp_r=0.64_Calph=-11.0_iangalp=11_p=-1.0_Comega=44000_ialp=3_mesh=061x101_pr=1.0_ivac=2_eta=0_NS=1_rot=MDI2_rho=1.butterfly_bp}}.
}
\end{figure}

\begin{figure}[!htb]
\centerline{\def\epsfsize#1#2{0.47#1}\epsffile{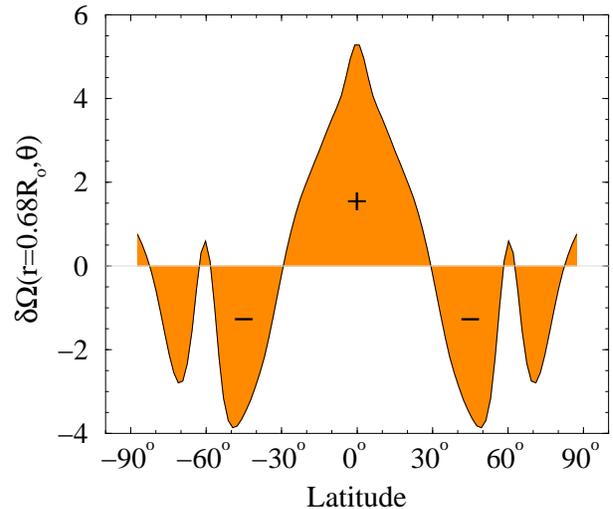}}
\caption{\label{opposite.signed.tachocline.r=0.68.MDI.sin4.data}
Variation of the perturbation to the zero order rotation rate in latitude and time,
revealing the
migrating banded zonal flows, after the
transients have died out.
Parameter values
are as in Fig.\ {\protect
\ref{dp_r=0.64_Calph=-11.0_iangalp=11_p=-1.0_Comega=44000_ialp=3_mesh=061x101_pr=1.0_ivac=2_eta=0_NS=1_rot=MDI2_rho=1.butterfly_bp}}.
}
\end{figure}

\item[(c)]
We observe the penetration of coherent torsional
oscillations into all the regions above the fragmentation level,
with the penetration extending to slightly greater depths in the case of
the GONG data.

\item[(d)]
Spatiotemporal fragmentation/bifurcation
can in principle explain, purely dynamically, the relationship between the
$11$ year and possible $1.3$ year oscillations near the
top and bottom of the convection zone.
In this connection, note that
$11=2^3\times 1.3$ years (compatible with the result
of Howe et al.\ 2000a being connected with three period halvings)!
It can also explain non-periodic dynamical
behaviour (compatible with the findings
of Antia \& Basu  2000).
\end{description}

Finally, apart from providing a possible theoretical framework for understanding
such phenomena, this scenario could, by demonstrating the
different qualitative dynamical regimes that can occur in the
dynamo models, also be of help in
devising strategies for future observations.

\begin{acknowledgements}
We would like to thank H.\ Antia, J.\ Brooke, K.\ Chitre, A.\ Kosovichev, R.\ Howe,
I.\ Roxburgh, M.\ Thompson, S. Vorontsov and N.\ Weiss for
useful discussions.
EC is supported by a PPARC fellowship.
RT would like to thank M.\ Reboucas
for his hospitality
during a visit to the CBPF, Rio de Janeiro,
where this work was completed.

\end{acknowledgements}

\end{document}